\documentclass[journal=jacsat,manuscript=article]{achemso}

\usepackage[version=3]{mhchem} 
\usepackage[english]{babel}
\usepackage[utf8]{inputenc}
\usepackage{graphicx}
\usepackage{SIunits}
\usepackage[nottoc]{tocbibind}
\usepackage{natbib}
\usepackage{xcolor}
\usepackage{caption}
\usepackage{tabularx}
\DeclareCaptionLabelFormat{adja-page}{\hrulefill\\#1 #2 \emph{(previous page)}}
\DeclareUnicodeCharacter{00A0}{~}

\author{Maryam Hosseini}
\affiliation[UNSW]
{School of Chemical Engineering, University of New South Wales, Sydney, NSW, Australia}

\author{Firoozeh Babayekhorasani}
\affiliation[UNSW]
{School of Chemical Engineering, University of New South Wales, Sydney, NSW, Australia}

\author{Ziyi Guo}
\affiliation[UNSW]
{School of Chemical Engineering, University of New South Wales, Sydney, NSW, Australia}

\author{Kang Liang}
\affiliation[UNSW]
{School of Chemical Engineering, University of New South Wales, Sydney, NSW, Australia}

\author{Vicki Chen}
\affiliation[UNSW]
{School of Chemical Engineering, University of Queensland, Queensland 4072, Australia}

\author{Patrick T. Spicer}
\affiliation[UNSW]
{School of Chemical Engineering, University of New South Wales, Sydney, NSW, Australia}

\email{p.spicer@unsw.edu.au }
\phone{+61 2 9385 5744}

\title[An \textsf{achemso} demo]
  {Propulsion, deformation, and confinement response of hollow nanocellulose millimotors}

\begin{document}





\begin{abstract}
\subsection*{Hypothesis:}
Micromotor and nanomotor particles are typically made using 
dense solid particles that can sediment or be trapped in confined 
flow environments. Creation of much larger motors should be possible 
if a very low-density system is used with sufficient strength to carry 
 liquid and still experience propulsive motion. Light, dense 
millimotors should also be able to deform 
more than dense solid ones in constrictions. 

\subsection*{Experiments:}
Millimotors are created from permeable capsules of bacterial cellulose 
that are coated with catalse-containing metal-organic 
frameworks, enabling 
reactive propulsion in aqueous hydrogen peroxide.
The motion of the motors is quantified using particle tracking and the 
deformation is measured using microcapillary compression and flow through 
confined channels.

\subsection*{Findings:}
Two different 
propulsion mechanisms are 
dominant depending on the motor 
surface chemistry:  
oxygen bubbles are 
expelled from hydrophilic 
millimotors, driving motion via 
reaction force and buoyancy. 
Hydrophobic 
millimotors remain 
attached to growing bubbles and move
by buoyancy alone. 
Despite their large 
size, the low-density capsules 
compress to 
pass through contractions 
that would impede 
and be blocked by 
solid motors.
The sparse structure but relatively 
large size of the motors enables 
them to transport significant volumes 
of liquid using minimal solid mass 
as a motor support structure.
\end{abstract}




\section*{Introduction}
Microorganisms can propel themselves through 
liquid by different swimming 
mechanisms, and synthetic 
particle motors, termed ``active matter'', 
have been created that mimic microbial 
motion by chemical, rather than mechanical, 
means \cite{moran2017phoretic,ma2021dual,tao2022nitric}.
Particulate nanomotors and 
micromotors move by converting 
chemical fuel from their environment 
into kinetic energy, and 
can achieve remarkable speeds 
relative to their body length 
\cite{gao2012catalytically,yan2021recent}.
Propulsion can be driven 
by self-generated solute 
gradients or electric 
fields \cite{moran2017phoretic} 
or by formation of gas bubbles 
that cause buoyancy or ejection-induced 
recoil effects \cite{wang2016nanomotors}.
Applications for the small motors 
are imagined in drug delivery 
\cite{tang2020enzyme,esteban2020multicompartment,gao2019superassembled,ma2021dual},
 environmental remediation 
\cite{zarei2018self,qiu2021interfacially,li2019step,liu2019bioinspired,ye2021magnetically}, and 
self-assembly \cite{vutukuri2020light,yuan2022reconfigurable} while 
the particles' unique motion is 
widely studied as well \cite{Marchetti2013}.
Larger, millimeter-scale motors 
have recently been 
developed from clay/DNA membranes 
to act as synthetic 
protocells that move and carry out internal 
biochemical reactions \cite{kumar2018enzyme}. 
These larger-scale motors 
can broaden the possible applications 
of active matter, motivating 
us to develop a 
 millimotor 
capsule that is easily functionalized 
but can also  
overcome difficulties most motors face 
with confined space navigation 
and sedimentation 
potential \cite{peng2018nanomotor}.

Our approach to develop these new 
millimotors takes inspiration 
from the biological cells 
that active matter seeks to 
mimic. Cells are partially permeable 
to water, minimizing density 
differences, while their softness 
enables them to deform 
and pass through narrow spaces and 
navigate confined environments.
By contrast, most synthetic 
active particles are 
made using dense solid materials, like 
platinum or silica, whose large 
density differences 
with water promote sedimentation. 
Such particles are also too rigid 
to deform and 
escape environmental 
confinement \cite{peng2018nanomotor}. 
A recent review \cite{xiao2018review} 
identified low 
density and robust deformability 
as important goals for future 
motor particles, and this work focuses 
on a new approach to meeting these challenges 
using unconventionally large motor capsules 
made from a mesh of bacterial cellulose fibers.

Hollow capsules are a 
promising way to minimise mass 
use and density issues in motor particles \cite{yang2020enzyme,terzopoulou2020metal,sanchez2010dynamics,zhu2018shapeable,ning2018geometry} 
and their low density makes them behave 
 like much smaller particles in 
fluid \cite{edwards1997large}. 
A recently developed bacterial cellulose 
capsule \cite{song2019soft} 
 provides a unique minimalist 
scaffolding for millimotor particle 
development, and we explore their motion and 
response after coating them with two different
MOF nanoparticles \cite{guo2019biocatalytic}.  
The MOFs attach onto the capsule's 
cellulose nanofibers and trap catalase 
enzyme in their structure, enabling 
conversion of aqueous hydrogen 
peroxide fuel into oxygen bubbles 
to drive 
motion \cite{kumar2018enzyme,sun2019enzyme}. 
The MOF surface polarity 
determines the mode of millimotor propulsion 
by altering oxygen bubble 
affinity for the capsule surface. 
The velocity and motion of 
the driven capsules 
are measured by 
optical microscopy and shown to be quite 
efficient compared to 
solid micro- and nanomotors, 
despite being much larger 
and full of liquid.
The low-density 
shells are efficient motor 
bases because of their structural integrity and 
minimal mass, but also provide unique 
benefits for bubble-driven flow. 
The flexible nanocellulose fiber struts allow 
significant deformation  
when passing through constrictions 
and the permeable capsules form a low-friction 
gas layer on the capsules that 
further enhances 
escape from confined spaces.

\section*{Methods and Materials}

\textbf{Cellulose microcapsule preparation:} 
Cellulose microcapsules with diameters in the 
range of \unit{50-1000}{\micro\meter} 
were synthesized by a biointerfacial polymerization 
process we previously developed
\cite{song2019soft}. In brief, bacterial 
cellulose microcapsules were grown using 
a water-in-oil emulsion of bacterial 
culture droplets as templates. 
The bacterial culture contains 
purified \textit{Acetobacter xylinum} 
concentrated from Kombucha culture 
(Nourishme Organics, Australia) by 
gradient centrifugation, coconut 
water (Cocobella, Indonesia), and 
10\% w/v table sugar. Within 10 days, 
the encapsulated bacteria polymerize 
glucose molecules into cellulose 
nanofibers, with a diameter 
of \unit{60-70}{\nano\meter}, 
that entangle to form a fiber mesh 
shell with 
a total thickness 
of \unit{20-50}{\micro\meter} 
and a pore size of 
\unit{0.5}{\micro\meter} \cite{song2019soft} 
at the oil-water interface. 
Subsequently, catalase-ZIFL and catalase-ZIF90 
MOF crystals were grown \textit{in situ} 
on the nanofibers, producing hydrophobic 
and hydrophilic 
millimotors, respectively. 
Capsules ranged between 
0.2-\unit{0.8}{\milli\meter} in size for both 
hydrophilic and hydrophobic particles.

\textbf{Hydrophobic ZIFL coating:} 
\unit{5}{\gram} of  
catalase from bovine liver, 
Sigma Aldrich), \unit{200}{\micro\liter} 
of  14.8 mM zinc nitrate ($ZnNO_3$, 
Sigma Aldrich) aqueous solution and \unit{2}{\milli\liter} of 714 mM  
2-methylimidazole (Sigma Aldrich) 
aqueous solution were added to 
\unit{1}{\milli\liter} of cellulose 
microcapsule dispersion. The 
mixture was mixed for 1 hr and 
then rinsed several times
with deionized water.

\textbf{Hydrophilic ZIF90 coating:} 
\unit{5}{\gram} of catalase, 
\unit{2}{\milli\liter} of  40 mM zinc 
nitrate ($ZnNO_3$, Sigma Aldrich) 
aqueous solution and \unit{2}{\milli\liter} 
of 160 mM  imidazolate-2-carboxyaldehyde (Sigma Aldrich) 
aqueous solution were added to 
\unit{1}{\milli\liter} of cellulose 
microcapsule dispersion. The 
mixture was mixed for \unit{1}{\hour} 
and rinsed with deionized water 
several times, Figure \ref{schematicmotors}.

\textbf{Enzyme labelling:}
\unit{8.5}{\milli\gram} of Rhodamine B 
isothiocyanate (RhB, Sigma Aldrich
Australia) was dissolved in 
\unit{0.5}{\milli\liter} dimethyl 
sulphoxide
(DMSO, Sigma Aldrich). In a glass 
vial, \unit{40}{\milli\gram} of catalase 
was placed in \unit{2}{\milli\liter} 
of sodium carbonate bicarbonate buffer 
(0.5 M, pH 9.5). Then, the RhB solution 
was added slowly into CAT solution. The 
CAT-RhB was then mixed for \unit{2}{\hour} 
at room temperature in darkness. The 
unreacted enzymes were separated from 
the labelled enzymes in an Illustra 
NAP-25 column (GE Healthcare). The 
first band eluted with Milli-Q water, 
which contains labeled enzymes, was 
collected for sample preparation.

\textbf{MOF and cellulose labelling:}
Congo Red and FITC were used to stain cellulose 
fibers and MOF crystals, respectivley to enhance 
microscopy visualization by 
addition of \unit{34}{\milli\gram} of  
0.5 wt\% aqueous Congo Red 
solution to \unit{1}{\milli\liter} 
of cellulose microcapsule 
dispersion. 

\textbf{Microscopy:}
Confocal and light sheet microscopy 
imaging experiments were carried out 
on a Zeiss LSM 880 with Airy 
scan \cite{macnab1972gradient}, and 
Zeiss Lightsheet Z.1 (Germany) microscope. 
A 63x oil immersion objective, with 
numerical aperture NA = 1.4, 5x dry 
objective with NA = 0.16, and 20x 
water immersion objective with NA = 1 were 
used depending on the scale of observation desired. 
Low-magnification 
optical microscopy 
images were acquired via stereoscope 
(WILD M3C, Leica, Germany) with 6.4x 
objective to enable individual 
particle tracking in 
microcapsule dispersions 
at room temperature. ImageJ 
software was utilized to quantify 
the fluorescence intensity inside 
and outside of the 
microcapsule \cite{Schindelin2012}.
Scanning electron microscope (SEM) 
images of the samples were taken on an FEI 
Nova Nano SEM 230 FE-SEM at an 
accelerating voltage of 5.0 kV.

\textbf{Fourier transform 
infrared spectroscopy (FTIR):} 
FTIR patterns were collected 
on Bruker IFS66/S High End 
FT-NIR/IR Spectrometer 
from 400 cm$^{-1}$ to 4000 cm$^{-1}$. 

\textbf{Microcapsule deformation:} 
Micropipette manipulation was 
used to apply controlled 
deformation to individual capsules. 
A microcapillary with a 
right-angle bend held the 
microcapsule in place 
while a second blunt microcapillary 
with an outer diameter of 1 mm 
was moved 
toward the microcapsule at 
a constant speed using a 
syringe pump stepper motor 
(Aladdin, WPI). 
The process was imaged at 
200 frames per second using an 
Opticam CMOS camera (Qimaging).

\textbf{Dissolved oxygen:} 
Dissolved measurements were performed 
using a dissolved oxygen meter 
(Oakton DO 6+)
to quantify propulsion 
reaction kinetics for both 
hydrophobic (ZIFL) and 
hydrophilic (ZIF90) motors 
in the presence of 
1\% v/v H$_2$O$_2$.

\section*{Results and discussion}
\subsection*{Capsule characterization}
\begin{figure*}
\centering
\includegraphics[scale=0.9]{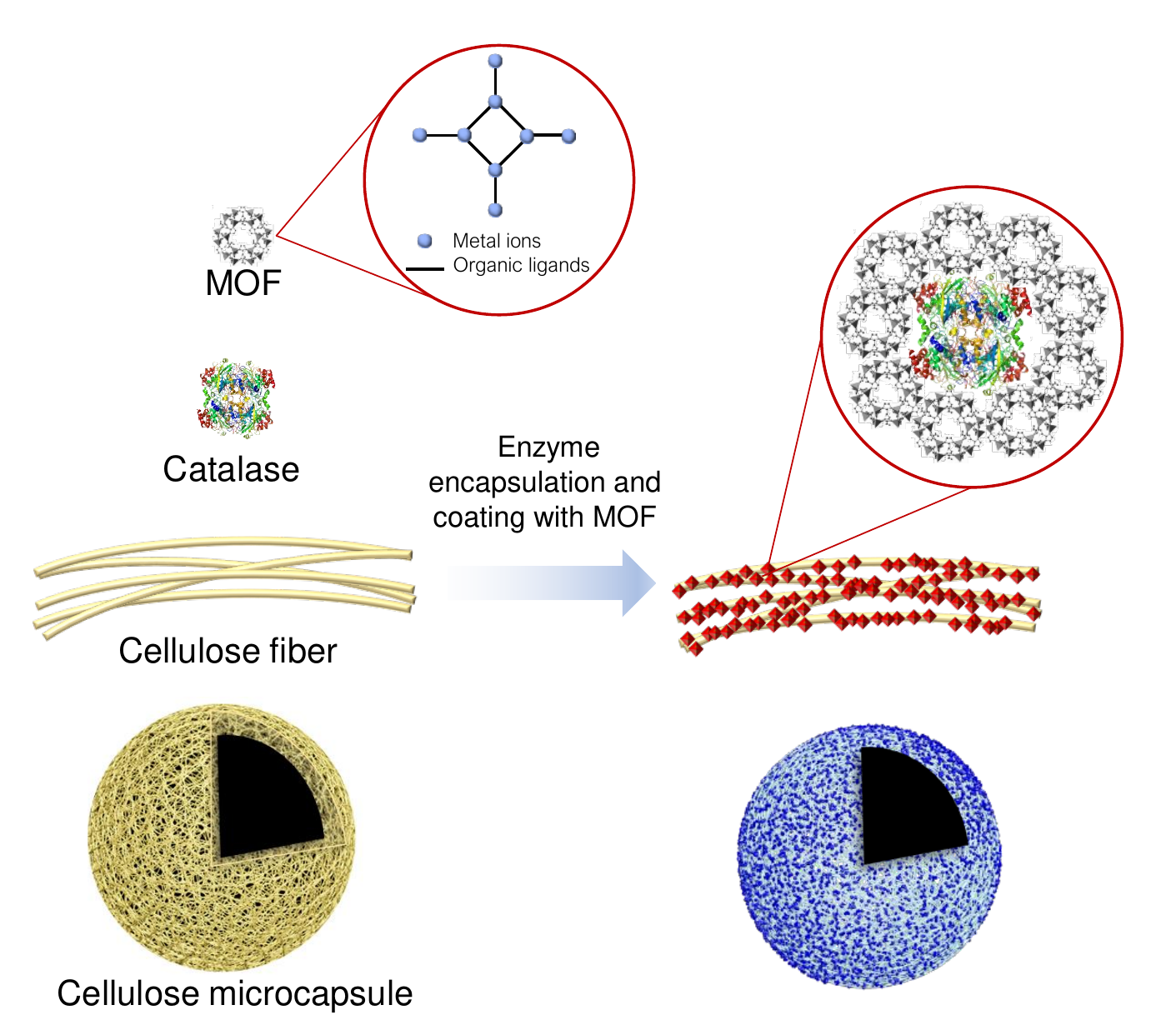}
\caption{Fabrication of hydrophilic and hydrophobic cellulose milllimotors. Schematic illustration of coating the bacterial cellulose microcapsule with MOFs that contain catalase and convert the capsules into milllimotors. The MOFs exist in the form of crystalline nanoparticles with a cage-like structure made of metal ions connected in a network via organic ligands and each crystal can encapsulate one or multiple biomolecules by physical adsorption \cite{liang2015biomimetic,chen2012can, liang2020biocatalytic}. The \textit{in situ} growth of MOF crystals on cellulose fibers occurs by adsorption of the positively charged metal ions onto the cellulose hydroxyl group through electrostatic interaction. Subsequent addition of organic ligands enables co-precipitation with the metal ions to form MOF crystals on the fibers \cite{ma2019multifunctional}.}
\label{schematicmotors}
\end{figure*}

Cellulose capsules are created using  
aqueous emulsion droplets of
\textit{Acetobacter} 
bacteria culture as templates.
The bacteria produce an entangled 
shell of cellulose fibers with micron-scale 
length and nanometer-scale 
thickness, Figure 
\ref{schematicmotors}.
The overall 
capsule diameter is millimeter-sized like 
the emulsion droplet templates 
used to grow them \cite{song2019soft}.
The use of the cellulose scaffolding 
provides a balance of structural integrity 
and flexibility with a signficant cargo 
volume.
For example, a \unit{0.5}{\milli\meter} diameter 
capsule with a \unit{20}{\micro\meter} shell thickness 
has a mass of only 
\unit{200}{\nano\gram} because of its 
high porosity, but its internal 
volume holds  400,000 times more water mass. 
The capsules are modified to enable 
fuel-driven motion by attaching 
a large number of active MOF 
nanoparticles to the fibers 
that contain catalase enzyme in 
their structure \cite{guo2019biocatalytic}.
The MOF particles created here 
are a crystalline matrix of zinc ions 
connected by two different organic ligands 
that allow us to produce motors coated 
with hydrophilic ZIF90 MOFs, using 
imidazolate-2-carboxyaldehyde ligand, 
and a version coated with hydrophobic 
ZIFL MOFs, using 2-methylimidazole
ligands, Figure \ref{schematicmotors}.
The 60 kDa catalase we used has a hydrodynamic diameter 
of $\sim$ \unit{7.4}{\nano\meter} and is encapsulated 
within the porous structure of the 
larger polycrystalline MOF nanoparticles 
that precipitate on the cellulose.
The process of coating cellulose nanofibers 
with MOF particles is shown schematically in 
Figure \ref{schematicmotors}. First, 
positively charged zinc 
ions are adsorbed on the cellulose hydroxyl 
surface group. After adding organic ligands, 
micron-scale MOF particles crystallize on 
the \unit{60}{\nano\meter} cellulose fibers, 
altering the capsule porosity and mechanical 
properties. The two MOF structures used have the 
same zinc metal ion basis but are connected 
by two different hydrophilic 
and hydrophobic ligands to 
vary the particles' surface chemistry and 
swimming behavior. 

\begin{figure*}
\centering
\includegraphics[scale=0.5]{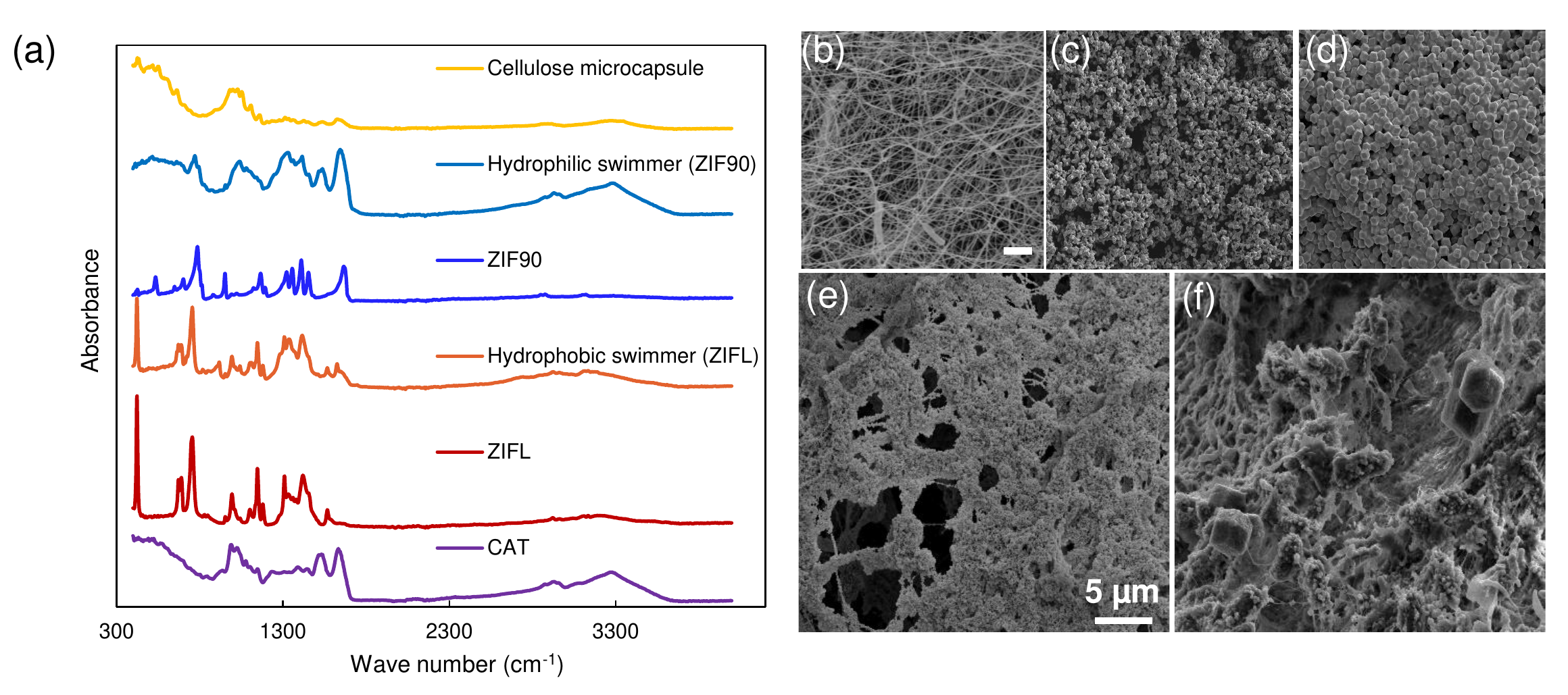}
\caption{(a) FTIR spectra of pure catalase (CAT), both MOF types, pristine cellulose microcapsules, and cellulose microcapsules after MOF coating. IR peaks show successful combination of enzyme, MOF, and cellulose fibers. (b) SEM of pristine cellulose microcapsule surfaces, (c) hydrophilic ZIF90 MOF particles, (d) hydrophobic ZIFL MOF particles. (e and f) Higher magnification views of cellulose fibers coated with MOF particles, highlighting the reinforcement of the cellulose microcapsule fibers by the dense deposits as well as some reduction in pore size. The MOF particle coatings on cellulose fibers tend to form pore-spanning bridges for (e) hydrophobic ZIFL MOF while deposits tend to follow the fiber structure more for (f) hydrophilic ZIF90 MOF. All scale bars \unit{5}{\micro\meter}.}
\label{Characterization}
\end{figure*}

FTIR was used to assess the success of MOF and enzyme coating on the microcapsules by measuring the presence of the ligand and enzyme chemical groups, Figure \ref{Characterization}.
For ZIFL, the characteristic peaks at 1585, 1147, $\sim$750 (double bonds) and 423 cm$^{-1}$ correspond to the stretching vibration of C=N, bending vibration of CH, bending vibration of the imidazole ring and vibration peak of Zn-N, 
respectively. The presence of these peaks indicates the 
significant presence of the 2-methylimidazole
ligands in the ZIFL.
The absorbance spectrum of ZIF90, 
shown in Figure \ref{Characterization}a), has a 
prominent mode centred at 1671 cm$^{-1}$, 
extending from 1751 to 1551 cm$^{-1}$, that 
is attributed to a C=O stretch \cite{lee2009surface} 
in the imidazolate-2-carboxyaldehyde ligand structure. 
The position and intensity of this carbonyl band 
obscures the amide I and, partially, the amide II 
spectral features of the catalase. 
Figure \ref{Characterization}(a) 
shows the amide II spectral region with two peaks 
centered at about 1540 and 1515 cm$^{-1}$. The 
position of these bands is consistent with the 
amide II components of free catalase, showing 
its successful encapsulation in the MOF particles 
on the cellulose fibers \cite{zhu2018shapeable}. 
Despite the 
lower sensitivity of amide II to protein secondary structure versus amide I, we conclude that the secondary structure of catalase-ZIF90 is comparable to catalase-ZIFL 
 \cite{zhu2018shapeable}. Structural studies of the 
 crystalline MOFs formed on the capsules showed 
 identical diffraction patterns for both ZIFL and 
 ZIF90 after enzyme encapsulation and attachment to 
 the cellulose capsule, Figure S1.

Figure \ref{Characterization}(b) shows an SEM 
image of native cellulose nanofibers in a capsule. 
The original cellulose fibers in the 
pristine capsules are quite strong and 
thin, but the deposition of metallic 
MOF nanoparticle networks onto these fibers 
will modify them in a number of ways, 
including their mechanical properties and 
the overall capsule permeability to the fluid environment.
The patterns of deposition will also 
affect the mode of propulsion by catalysis, 
so we characterize their state by imaging.
Figures \ref{Characterization}(c-f) show crystals 
of ZIFL and ZIF90 deposits on the capsule fibers. 
The microcapsules were 
freeze-dried at \unit{-65}{\celsius} 
 to avoid any collapse that might occur 
by capillary pressure during air drying. 
As shown in Figure \ref{Characterization}, 
the fiber diameter is between 20-70 nm with 
a length of several microns for the unmodified 
cellulose microcapsule \cite{song2019soft}. 
After crystallization of MOF on the capsules, 
particles with an average size of \unit{100-150}{\nano\meter} for 
ZIFL and \unit{150}{\nano\meter} for ZIF90 attached to the 
surface of the cellulose fibers, 
Figure \ref{Characterization}(c-d). 
The addition of the MOF particles to 
the capsule is expected to alter the mass and 
inertia of the system during subsequent propulsion 
studies.
If we assume the MOF coating is a 99\% dense single 
layer of \unit{100}{\nano\meter} particles with a density 
of \unit{2}{\gram\per\centi\meter^3}, the 
particles add \unit{150}{\nano\gram} mass to 
a \unit{0.5}{\milli\meter} diameter cellulose capsule, 
at most doubling its weight 
while still being dominated by the 
mass of the liquid cargo in its capsule.

Differences can be seen in the coating 
morphology at higher magnifications, 
Figure \ref{Characterization}(e-f), based on 
the polarity of the MOF particles produced.
For example, the hydrophobic ZIFL particles 
form structures that partially 
span the gaps between fibers
Figure \ref{Characterization}(e), 
likely because the 
particles have less affinity 
for the hydrophilic cellulose.
The hydrophilic ZIF90 coatings seem to follow 
the fiber structures more 
closely due to more intimate contact, 
Figure \ref{Characterization}(f).
Although different morpholgies form on 
the capsules, MOF film growth 
on individual fibers 
will in either case increase 
the effective fiber size and 
reduce the pore size of the overall capsule. 
Previous measurements \cite{song2019soft} 
indicated the pristine capsules 
have pore sizes on the order of \unit{500}{\nano\meter} 
so the MOF crystals will 
rapidly reduce the overall capsule 
permeability as they grow.

\begin{figure*}
\centering
\includegraphics[scale=0.8]{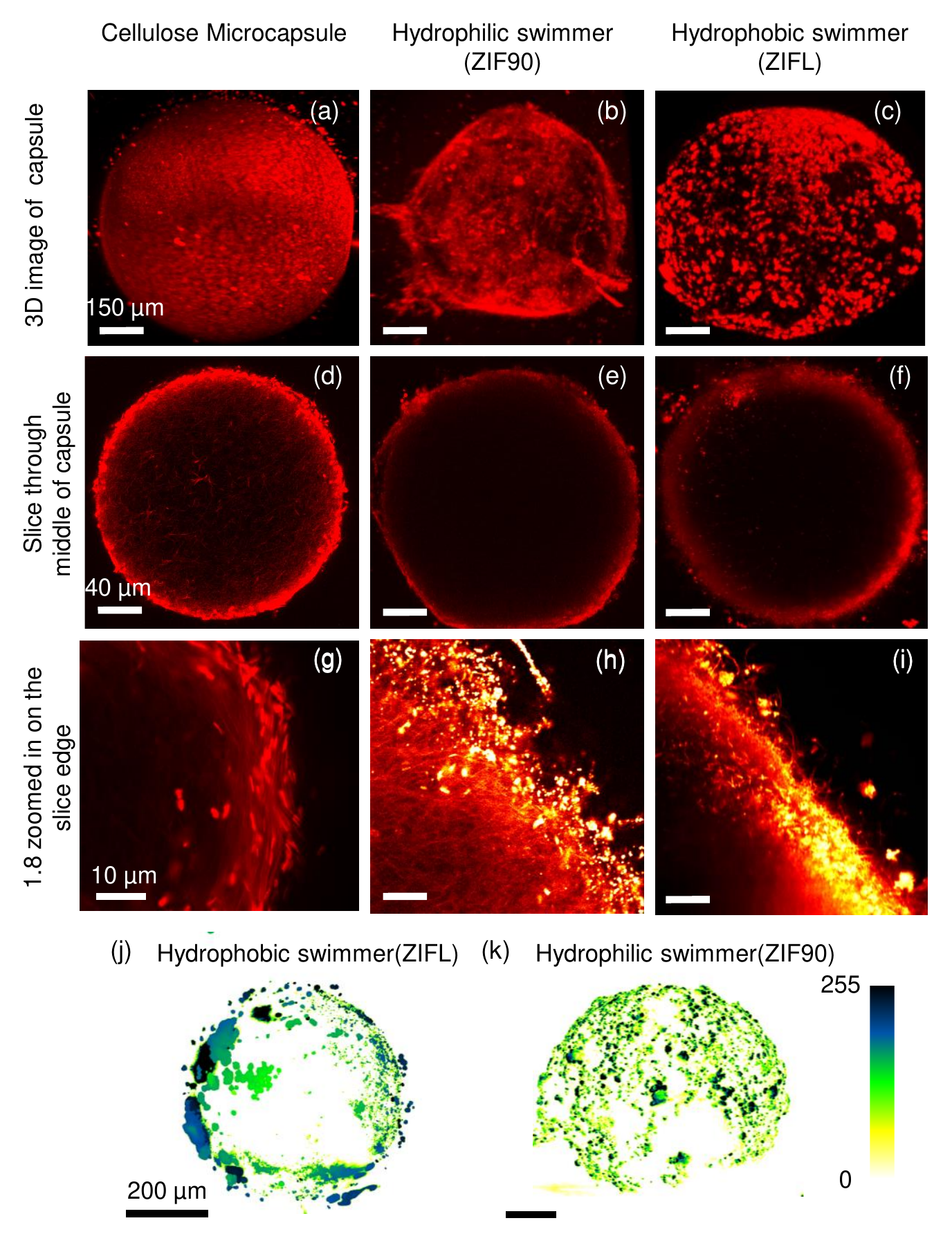}
\caption{Lightsheet fluorescence microscopy shows 3D reconstructions of the (a) pristine cellulose microcapsule, (b) a microcapsule coated with hydrophilic ZIF90, and (c) hydrophobic ZIFL MOFs. A single mid-plane slice through the microcapsule (d) before and after coating with (e) hydrophilic and (f) hydrophobic MOF. (h-i) Higher magnification confocal images show MOF particles in yellow and cellulose fibers in red. 3D images of enzyme distribution on the (j) hydrophobic and (k) hydrophilic millimotor surface.}
\label{Confocal microscopy}
\end{figure*}

Different microscopy techniques 
were used to visualize the pristine 
and modified cellulose microcapsule 
at larger length scales for 
a more holistic capsule characterization. 
Light sheet fluorescence microscopy 
(LSFM) allows us to visualize the full 3D 
structure of the millimotors. Cellulose, 
MOFs, and catalase were stained with 
Congo Red, FITC, and Rhodamine B, 
respectively, before fluorescent imaging. 
Figures \ref{Confocal microscopy}(a-c) 
show 3D reconstructions of 
pristine and modified cellulose 
microcapsules coated by both MOF types, 
with the cellulose fibers shown in red.
At this magnification, the pristine capsule 
in Figure \ref{Confocal microscopy}(a) looks 
spherical and almost solid as 
the capsule pores are 
micron-scale \cite{song2019soft}.
A capsule with the hydrophilic ZIF90 MOF 
coating is shown in 
Figure \ref{Confocal microscopy}(b) and the 
impact of the MOF coating is clearest in the 
dark regions of the spherical capsule where larger 
MOF regions obscure the cellulose fibers. 
The coating is heterogeneous due to random 
variations in the solution during precipitation.
More detail can be resolved at higher magnifications.

Higher resolution imaging is performed 
using confocal microscopy to visualize the 
hollow and fibrous structure of the 
microcapsules, with and without a MOF coating, 
Figures \ref{Confocal microscopy}(d-i). 
Figures \ref{Confocal microscopy}(d-f) 
show a single mid-plane slice 
through the capsule, 
demonstrating that the 
capsules are largely hollow, 
and have a shell on the order of 
\unit{20}{\micro\meter} in agreement with
past work \cite{song2019soft}. 
The capsules with hydrophilic ZIF90 and 
hydrophobic ZIFL MOF coatings in Figures 
\ref{Confocal microscopy}(e-f) have similar 
shell thicknesses to the pristine capsule in 
Figure \ref{Confocal microscopy}(d), 
as the nanoscale MOF 
coatings coat individual fibers and 
don't significantly 
change the overall capsule wall thickness.
An even higher magnification view of the capsule 
shells highlights the degree of penetration of 
the fiber structures by the MOF particles 
being deposited, 
Figures \ref{Confocal microscopy}(g-i). 
Here the MOF nanoparticles 
are shown as yellow and 
the cellulose fibers as red.
The optical resolution of the 
confocal system used here 
is on the order of \unit{200}{\nano\meter} so these 
images will not resolve the 
smallest deposits formed and only 
serve here to indicate the 
degree of individual fiber 
coating by the MOFs.
As the MOF particles are 
the framework we use to encapsulate the 
catalase enzyme for propulsion, the distribution 
of both is of interest to particle design and 
motor performance.

Enzyme distribution is a key parameter 
determining millimotor motion and 
orientation \cite{patino2018fundamental}. 
The distribution of enzyme on the 
MOF-coated capsules is demonstrated in 
Figures  \ref{Confocal microscopy}(j-k) 
using a 3D reconstruction of the millimotor 
capsules with a color map indicating the 
intensity of the
Rhodamine B dye labeling the catalase.
We observe a heterogeneous but thorough 
distribution of the enzyme on the 
surface of the hydrophobic motor, 
Figure \ref{Confocal microscopy}(j), 
but an asymmetric distribution along 
the top half of the surface of the 
hydrophilic millimotor, possibly 
because of a higher affinity of the 
enzyme for the hydrophobic MOFs, 
causing some segregation on the 
hydrophilic system \cite{liang2019enhanced}.
A bulk-scale measurement of enzyme 
availability, a Bradford 
assay \cite{bradford1976rapid}, 
was also used to calculate the 
encapsulation efficiency of catalase 
as 98.2\% for hydrophobic ZIFL and 
85.9\% in hydrophilic ZIF90 motors. 
The difference in spatial distribution 
of the enzyme on the 
capsules could potentially  
affect self-propulsion of 
the motors by inhomogeneous oxygen 
bubble production \cite{patino2018fundamental,patino2018influence} 
but for buoyancy-dominated motion such an 
effect should be negligible.
The MOF coatings on the capsules can also 
affect porosity of the capsules, 
so we now assess modified capsule 
exchange with the fluid environment.

\begin{figure*}
\centering
\includegraphics[scale=0.7]{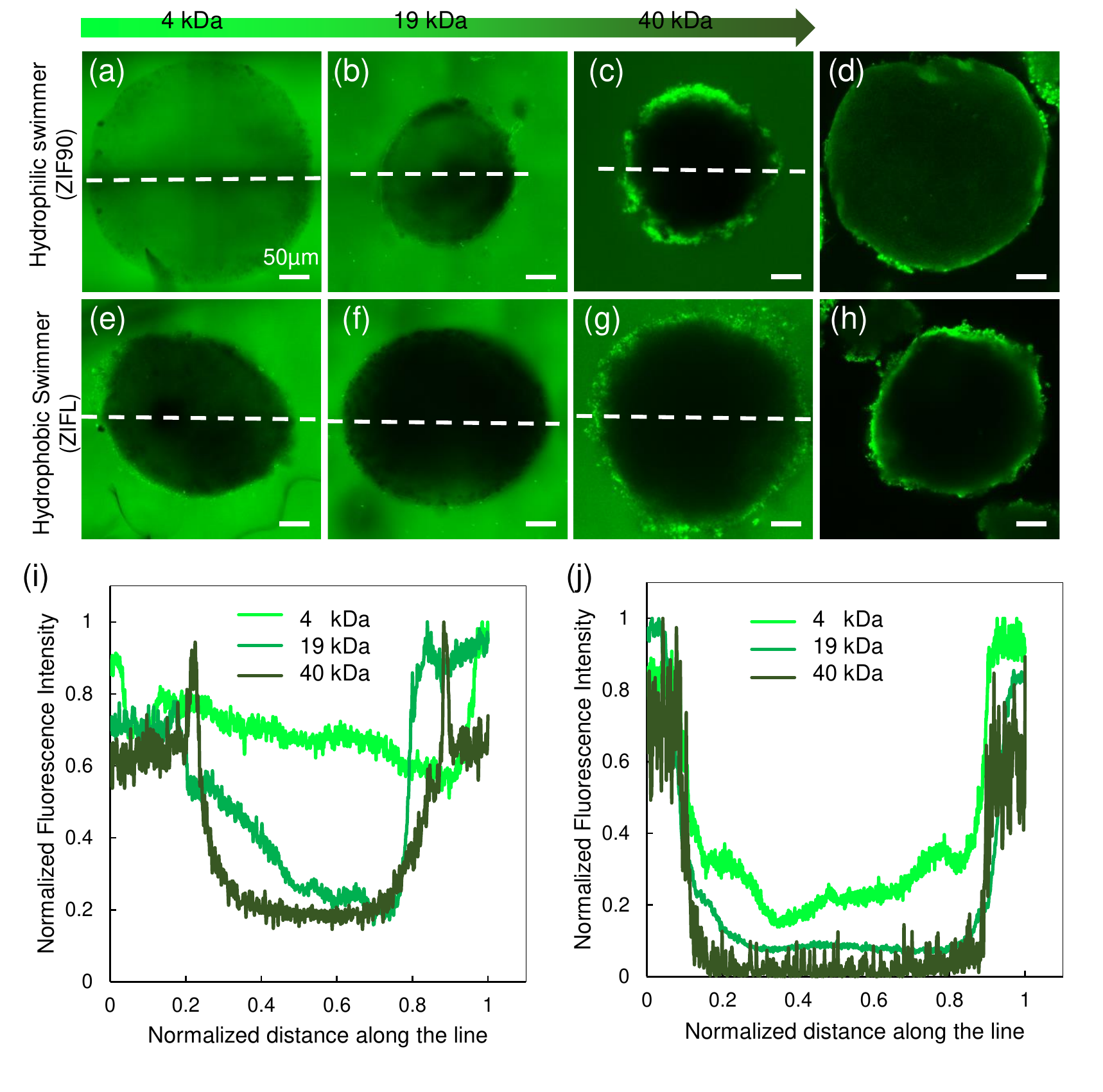}
\caption{Measurement of permeability of (a-c) hydrophilic and (e-g) hydrophobic millimotors based on transport of three different FITC-dextran tracers with molecular weight ranging from 4-40 kDa. Figures (d) and (h) show capsules that contained 40 kDa dextran before being coated with MOFs to demonstrate their encapsulation ability.  The FITC-dextran concentration varies from green regions of high concentrations to black where no FITC-dextran is resolved. Scale bars are \unit{50}{\micro\meter}. The fluorescence intensity in (a-g) is summarized by plotting integrated normalized intensity of the images for the (i) hydrophilic ZIF90 and (j) hydrophobic  millimotors.}
\label{Permeability}
\end{figure*}

\subsection*{Porosity of motors}
Song et al. \cite{song2019soft} studied the 
permeability of pristine cellulose microcapsules 
and found a pore size on the order of \unit{500}{\nano\meter}. 
After surface modification 
of the microcapsules by MOF deposition, 
Figure \ref{Confocal microscopy}, 
we expect the pore size 
to change significantly 
so we study this by molecular 
tracer diffusion experiments.
FITC-labeled dextran chains ranging from 
4-40 kDa were used as tracers to determine 
the permeability of MOF-coated cellulose 
microcapsules based on size exclusion. 
After mixing microcapsules with 1 mg/ml 
of FITC-dextran and 24 hr incubation, 
confocal microscopy images of capsules 
were obtained to determine 
the ability of different sizes to 
penetrate the coated capsules. 
Figures \ref{Permeability}(a-c) 
show capsules that have been exposed 
to different size FITC-dextrans and the 
intensity of green tracer that was 
able to diffuse inside of the 
capsules coated with hydrophilic ZIF90 
provides a measure of 
the new capsule pore size.
The 4 and 19 kDa tracers penetrate the 
hydrophilic ZIF90 microcapsules in 
Figures \ref{Permeability}(a-b), 
indicating a pore size larger 
than \unit{1.9}{\nano\meter},
while the 40 kDa remains 
outside of the capsule, 
Figure \ref{Permeability}(c), blocked 
by the new pore size and indicating 
the hydrophilic capsule is 
impermeable to \unit{4.3}{\nano\meter} species \cite{armstrong2004hydrodynamic}.

For hydrophobic ZIFL 
coated capsules, 
4 kDa FITC-dextran shows a small 
amount of diffusion inside the 
capsule after \unit{24}{\hour}. 
However, 19 and 40 kDa molecular 
weight FITC-dextran chains  
are totally excluded from the microcapsules. 
These capsule pores must then have an average diameter  
less than \unit{1.9}{\nano\meter} \cite{armstrong2004hydrodynamic}. 
A second experiment was performed to 
confirm the upper limit of pore size 
of both capsule types: we 
immersed microcapsules in 
40 kDa FITC-dextran and \textit{then} 
coated them with hydrophilic 
and hydrophobic MOFs. 
After washing the capsules with deionized 
water multiple times, and holding for \unit{24}{\hour}, 
we saw no release of FITC-dextran from either 
hydrophilic or hydrophobic capsules, 
Figures \ref{Permeability}(d) 
and (h), respectively. 
Different reaction times and conditions 
could alter these effects, but the results in Figure \ref{Permeability} indicate 
a more than 100X 
variation in pore size is possible 
for MOF-caoted cellulose capsules. 
For comparison, plant cells have a 
cellulose structure that is permeable 
to molecules with a diameter 
ranging from \unit{3.5-5.2}{\nano\meter}  \cite{carpita1979determination}.
MOF coating of capsule fibers enables 
us to create motors with chemical propulsion 
that have tunable permeability and solute 
exchange.

Because the dextran is shown to be uniformly 
blocked by the coating, it indicates a fairly 
uniform coating, on average, by the crystals 
given the deposition 
occurs throughout the 
thickness of the 
capsule walls.  
So these particles 
remain permeable after coating, 
though much less so than their 
more permeable pristine form, 
meaning they can fill with fluid 
and carry that cargo during movement. 
It also means the coated 
capsule millimotors can  
still exchange contents 
with the environment 
via molecular diffusion, 
for example to perform 
delivery of a chemical 
cargo \cite{kumar2018enzyme}.
The length scale of the pores can then 
be used to control the 
subsequent selectivity 
of the capsule 
in these applications.

 \begin{figure*}
\centering
\includegraphics[scale=0.47]{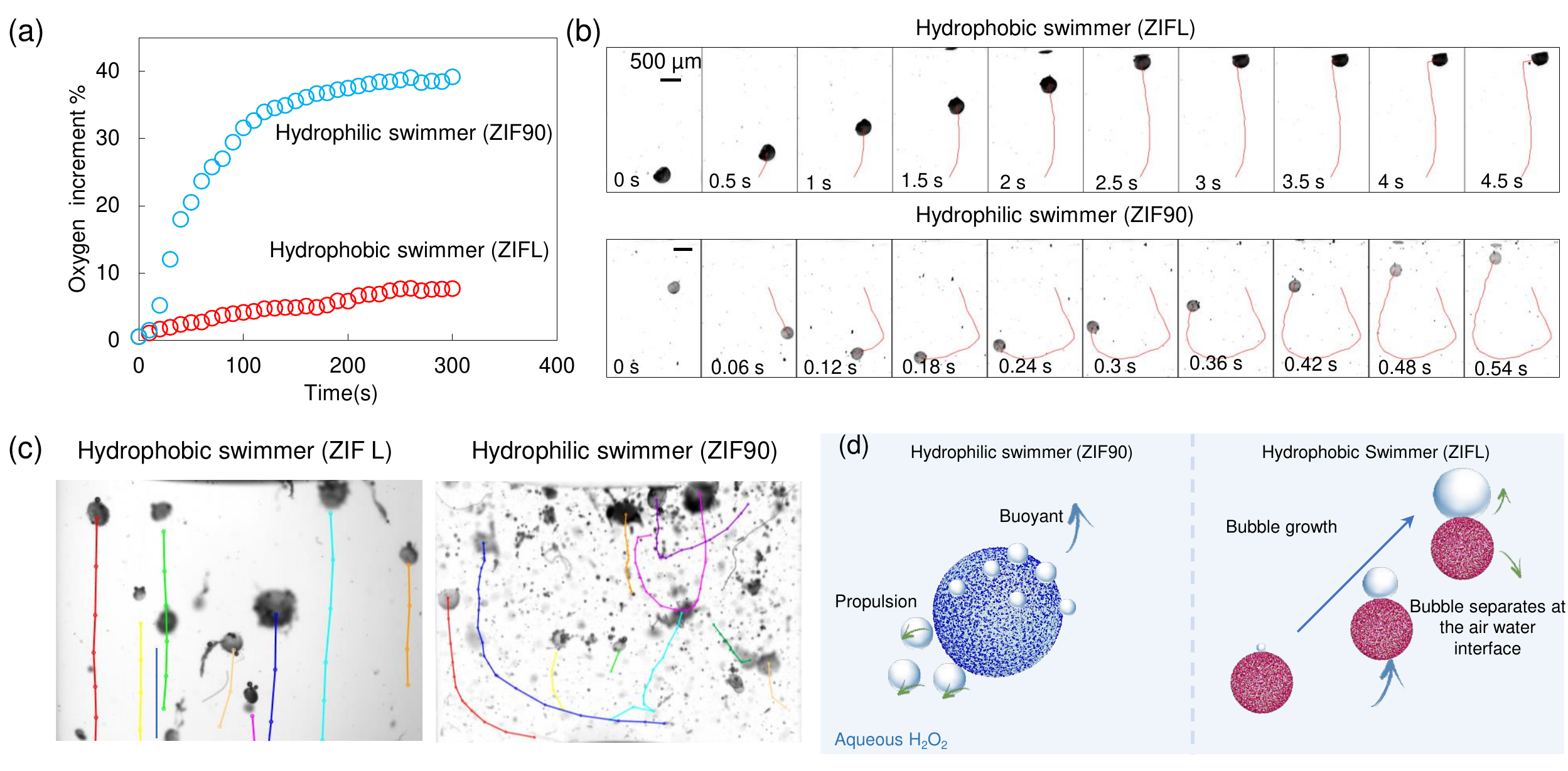}
\caption{ (a) Oxygen gas increment measurements in an aqueous 1\% hydrogen peroxide solution containing dispersed hydrophilic or hydrophobic millimotors. (b) Time-lapse images of hydrophobic ZIFL (first row) and hydrophilic ZIF90 (second row) motor trajectories in aqueous 0.25 wt\% H$_2$O$_2$. For hydrophobic motors, bubble formation and growth lifts motors vertically via buoyancy.  For hydrophilic motors, both bubble propulsion and buoyancy contribute to motions, initially driving motors in random directions then mostly vertically. (c) Multiple hydrophilic and hydrophobic motors have similar traits, moving vertically when buoyancy dominates and in multiple directions when both buoyancy and propulsive recoil contribute.  (d) Schematic of the proposed mechanism for millmotor bubble formation and movement: hydrophobic motors remain attached to O$_2$ bubbles and their motion is dominated by buoyancy, while hydrophilic motors have elements of horizontal motion due to propulsion by bubble ejection as well as buoyancy.}
\label{bubbleratemechanism}
\end{figure*}

\subsection*{Millimotor capsule propulsion}
Oxygen gas production and bubble 
formation play an essential role in 
propulsion of both hydrophobic and 
hydrophilic millimotors. When 
exposed to an aqueous 
hydrogen peroxide solution, catalase 
enzyme immobilized in the MOFs 
catalyzes hydrogen peroxide 
decomposition into oxygen and water 
molecules, with the excess oxygen gas 
forming bubbles and driving motion.
We first assess the rate of reactive 
oxygen generation in a bulk solution 
of 1\% v/v\% H$_2$O$_2$ containing 
hydrophobic ZIFL and hydrophilic 
ZIF90 motors, Figure 
\ref{bubbleratemechanism}(a). 
Both systems rapidly produce oxygen, but 
the rate of O$_2$ production by 
hydrophilic ZIF90 motors was 
more than six times 
that by hydrophobic ZIFL motors, 
Figure \ref{bubbleratemechanism}(a),
as a result of differences in enzyme 
activity. The catalase enzyme has 
a higher affinity for a hydrophobic 
surface, which can cause conformation 
changes that denature the protein, 
reducing its activity \cite{liang2019enhanced}.
Oxygen production rate is important to motor 
performance, as sufficiently high rates 
can lead to bubble ejection and propulsion 
by recoil forces, 
whereas slower growth tends to favor bubbles 
remaining attached to a motor so that motion 
is dominated by buoyancy.
The mode of bubble-motor interaction, 
however, is also affected by the motor surface 
chemistry and the resultant 
affinity of  
hydrophobic bubbles for a motor.

We studied the performance 
of these motors by observation of 
an aqueous dispersion of 
individual or multiple 
enzyme-powered millimotors.
The motors were placed in 
a transparent 
cuvette and allowed to sediment, 
then different concentrations of 
aqueous hydrogen peroxide 
solution ranging  
from 0.065 - 1\% were added 
and allowed to diffuse to 
the motors at the bottom to 
initiate propulsion. 
Once peroxide decomposition began, 
cellulose millimotors became 
buoyant and migrated from a 
resting position at the bottom 
of the cuvette to the rest of the 
fluid volume.
A high-speed camera was 
utilized to observe growth of 
single and multiple 
oxygen bubbles inside or 
outside of the motors
in less than a minute after 
addition of hydrogen peroxide.

In all motion studies, we see a clear 
difference between the hydrophobic 
ZIFL motors, that initially move 
in all directions 
as a result of rapid 
bubble production 
and ejection, 
and the hydrophilic ZIF90 motors that only 
move vertically, 
Figure \ref{bubbleratemechanism}(b).
For example, the hydrophobic 
ZIFL motor
in the top row of 
Figure \ref{bubbleratemechanism}(b)
follows a completely vertical trajectory 
over several seconds until it reaches the top 
of the water volume and drifts slightly 
to the side.
This is consistent with extensive work on 
mineral flotation, where hydrophobic particles 
are seprated from suspensions using air bubbles 
that flow through the 
system \cite{van1999gas,vinke1991particle}.
The hydrophilic ZIF90 motor in the bottom 
row of Figure \ref{bubbleratemechanism}(b), 
however, initially moves down as a result of 
ejecting a bubble upward, then makes a hard turn 
to our left and moving across the bottom of 
the cuvette.
After these two major direction changes, the 
motor then moves up as buoyancy becomes 
dominant.
Multiple millimotors in dispersion 
behave similarly, as shown in 
Figure \ref{bubbleratemechanism}(c) 
for a 0.05 wt\% solution of 
hydrogen peroxide.
The hydrophobic 
ZIFL motors all follow remarkably 
consistent vertical trajectories as 
their motion is dominated by 
buoyancy due to growth of 
strongly attached bubbles.
The hydrophilic ZIF90 motors, however, show more 
complex motion that can lead to 
correlated trajectories, like the two motors in 
the bottom left corner of the image, but 
opposing directions are also possible as seen 
in the upper right region of the image.
The enzymatic activity is maintained until the 
exhaustion of the chemical fuel, so adjusting 
the concentration increases the initial rates 
of movement and extends the length of time that 
movement can be observed.
The two motor surface chemistries 
thus 
enable different directions of movement. 
Bubbles can be generated 
wherever enzymes are 
located on the surface 
of the motors, so if the 
bubble is initially 
ejected the motor will move 
in the opposite direction by recoil.
If bubbles remain attached, however, the 
motor will move up by buoyancy and its 
initial orientation is less significant to 
the subsequent motion.
We summarize the mechanisms of motion  
by schematic in 
Figure \ref{bubbleratemechanism}(d).

The cellulose microcapsules used 
here vary in size 
from \unit{200-600}{\micro\meter}, 
all much larger than commonly-studied micro- 
and nanomotors, but consistent 
with recently-developed protocell 
millimotors \cite{kumar2018enzyme}.
We are curious about the performance of 
these motors versus 
their smaller counterparts, however, 
as their large size but low solid mass could 
provide additional benefits.
As noted earlier, the capsules are low-density 
and permeable, with the 
majority of their inertia 
due to water that permeates the interior.
A useful benchmark is the motor velocity 
relative to its diameter, where impressive 
values of 4-200 
diameters/second are 
known \cite{gao2012catalytically}.
Here we see hydrophobic ZIFL  
millimotors regularly moving at 
average velocities of up 
to \unit{4}{\milli\meter\per\second}, 
a relative velocity of more than 8 
diameters/second, while 
hydrophilic ZIF90 
millimotors can reach average 
velocities of \unit{40}{\milli\meter\per\second}, 
a relative velocity of more than 80 
diameters/second. 
The difference is the result of the 
recoil force exhibited by only the 
hydrophilic ZIF90 motors, as buoyancy is 
the only other mechanism 
possible for both systems.
Given that these motors are orders of 
magnitude larger than micro- or 
nanomotors, it is encouraging 
to note their 
propulsion, even while containing 
a massive liquid cargo,
is competitive with much smaller solid 
motors.

\begin{figure*}
\centering
\includegraphics[scale=0.7]{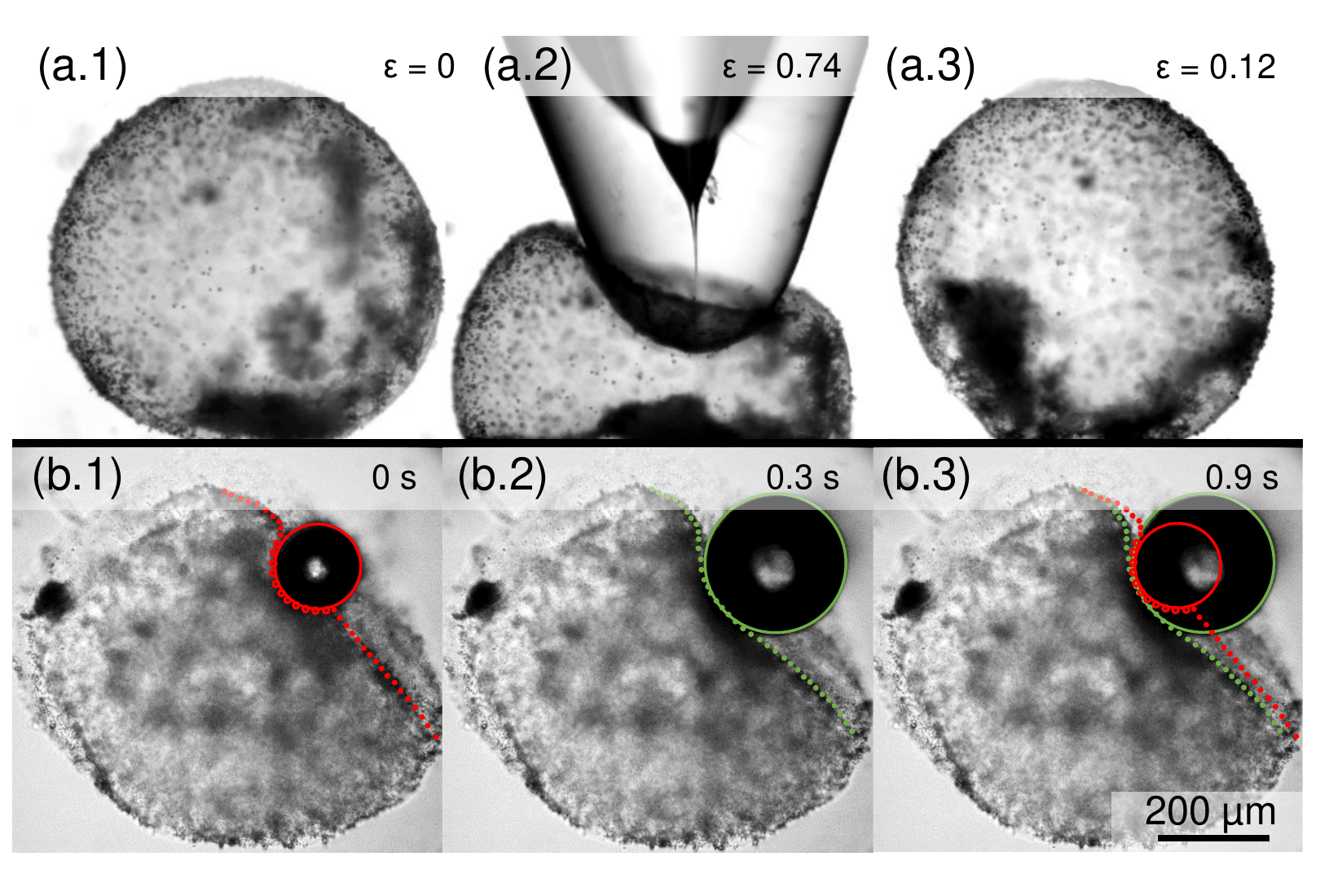}
\caption{Deformation and recovery of hydrophobic (a1-a3) motors using
microcapillary manipulation. The microcapsule was deformed by a capillary tip, and each column indicates the stage of initial, deformation, and recovery.
The spherical shape of  the hydrophobic motor is recovered after 75 \% of strain. (b1-b3) Successive images show the growth of an oxygen bubble on the surface of a hydrophobic ZIFL capsule over 5 millisecond intervals. The capsule deforms under the stress exterted by the bubble, indicating an ability to change shape during movement 
as the drawn lines document.}
\label{capillarydeformation}
\end{figure*}

\begin{figure*}
\centering
\includegraphics[scale=0.65]{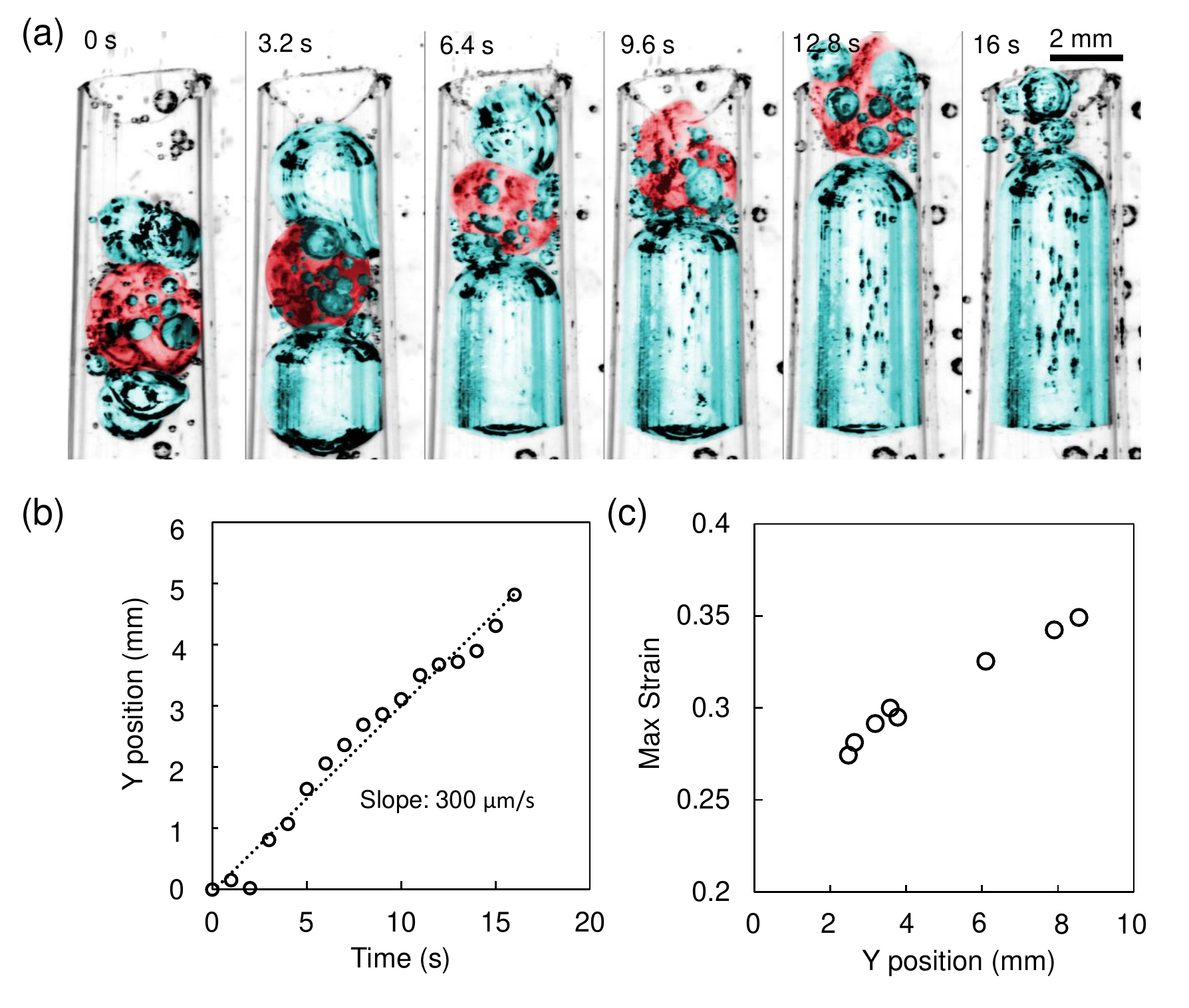}
\caption{(a) Composite images of the hydrophilic motor passing through a constriction in an aqueous solution of 3\% H$_2$O$_2$. (b) The slope of the vertical position and time plot shows the velocity of the hydrophilic motor, around \unit{300}{\micro\meter\per\second}, during its motion through the confined channel. (c) Changing hydrophilic motor strain at different vertical positions in the constriction.}
\label{confinementpropulsion}
\end{figure*}

\subsection*{Motor deformation and confinement response}
The cellulose motors have 
a unique porosity and response 
to mechanical stress 
because of their combined 
strength and flexibility \cite{song2019soft}.
As a result, we are interested 
in how these particles affect 
capsule response to 
extensional stress. First, we 
need to evaluate the extent to 
which the modification of the 
motors by the formation of MOF 
crystals affects 
their mechanical properties.
Song et al. \cite{song2019soft} 
found unmodified capsules 
deformed elastically under compression, 
recovering after deformations 
as large as 20\% but remaining 
indented or buckled at higher 
strains.
Adding crystalline MOF particles, 
however, is 
expected to stiffen 
the fibers and 
change the capsule 
response to deformation. 
Stress-strain curves for 
up to five individual millimotors 
were measured using an 
extensional flow in a 
microfluidic channel \cite{song2019soft}. 
We found that the elastic 
modulus, $E=$\unit{100}{\pascal}, 
of pristine cellulose 
microcapsules increased 
$\sim 30X$ after coating 
with hydrophilic ZIF90 
MOFs and $\sim 140X$ 
after coating with 
hydrophobic ZIFL MOFs. 
The difference is
likely a result of variations in 
coating 
uniformity and 
distribution, Figure \ref{Characterization}. 

Along with an increased elastic modulus,
the capsules preserve significant 
ability to recover from 
deformation, as we 
demonstrate in 
Figure \ref{capillarydeformation}.
Figure \ref{capillarydeformation}(a) 
shows the microscopic response 
of a hydrophobic ZIFL 
millimotor to deformation 
by a blunt microcapillary end. 
Both hydrophobic ZIFL and 
hydrophilic ZIF90 motors largely recover their 
original profile after being 
indented by external stress 
to a strain of $\varepsilon = $74\% 
and higher, while
pristine capsules do not fully 
recover from strains higher than 
$\varepsilon = $20\% \cite{song2019soft}. 
The addition of MOFs clearly increases 
the elasticity of the capsules 
while adding propulsive capability. 
This is consistent with above 
estimates of a significantly increased  
capsule modulus after MOF coating.

Figure \ref{capillarydeformation}(b) 
suggests an even more interesting 
dimension to the flexibility of 
these motors. A series of successive 
close-up images of a 
bubble evolving on a hydrophobic ZIFL 
motor surface in 
Figure \ref{capillarydeformation}(b)
indicates that the motor structure 
actually flexes in response to 
the pressure of the bubble 
expansion itself. 
Here the bubble is likely 
generated inside of the 
capsule and then squeezes 
out of a pore to 
exert the pressure seen here.
Such an effect could also 
occur if the bubble formed 
on a capsule that was confined in some way.
The result indicates the motors 
could actually be changing shape 
during motion, increasing their 
flexibility and responsiveness, 
and offering a mechanism 
for more complex 
movements in future work.
As pointed out earlier, there is a 
need for motors to navigate confined 
environments while adapting and 
recovering to the different 
conditions they encounter. 
We now evaluate whether such 
deformation and response 
can occur when the millimotors 
encounter confinement during their 
propelled movement.
We do this by studying 
the performance of the 
motors during propulsion in a 
constricted channel and note whether the 
combination of propulsion 
and flexibility enhance their mobility.

Figure \ref{confinementpropulsion}(a)
shows a series of images of a 
hydrophilic ZIF90 motor as it moves 
vertically through a gradually 
constricting capillary. 
The spherical millimotor initially 
has a diameter of \unit{2.9}{\milli\meter} 
so can enter the capillary but must compress 
in order to exit, as the capillary internal 
diameter reduces from 
\unit{3.5}{\milli\meter}, 
to \unit{2.2}{\milli\meter} 
at its outlet. 
The motor has been artificially 
colored red in the image 
and the surrounding 
bubbles have been colored blue 
to enhance visibility.
The motor is initially 
driven by buoyancy 
but slows in the second 
frame in response to the narrowed passage.
The narrowing of the channel 
causes compression of the motor, 
increasing its strain to more 
than 25\%, Figure \ref{confinementpropulsion}(c). 
However, the motor continues to 
produce bubbles behind itself 
that grow as the 
reaction proceeds. 
A solid particle at this point 
would be unable to proceed and 
simply block the channel, 
but the flexible motor is able to deform 
much more in response to stress, 
Figure \ref{capillarydeformation}(a), 
and continues to move, albeit 
in a more cyclical way as seen 
in the plot of vertical position in 
Figure \ref{confinementpropulsion}(b). 
Despite the small variations 
in velocity, the overall progress of 
the motor is relatively constant as 
a result of its flexibility and 
reactive propulsion.
The growth of surface bubbles likely 
aids movement by reducing drag 
on the capsule. 
Figure \ref{confinementpropulsion}(a) 
also shows the formation 
of a large bubble 
behind the motor that 
aids in pushing it 
through the constriction. 
Interestingly, the bubble driving 
the capsules through the capillary 
is pinned to the capillary wall, 
allowing it to push against the 
capsule. Such behavior is 
beneficial for this additional 
propulsion mechanism to act, as a 
non-wetting capillary surface would 
reduce the ability of 
the bubble to push the motor. 
The results in 
Figure \ref{confinementpropulsion} 
show that the production 
of deformable responsive motors 
with the low-density 
cellulose 
capsules developed here enable 
use in confined 
environments that solid 
particle counterparts can 
not handle. A high level of deformation 
might be expected to 
cause erosion of the MOFs from the 
capsules surface; 
however, no detectable fragments were 
observed following the deformation experiments 
in Figure \ref{capillarydeformation}(a).
In addition, the results in Figure \ref{confinementpropulsion} 
indicate propulsion is maintained even during 
quite robust deformation in 
confinement, further confirming the 
motors largely remain attached.

\section*{Conclusions}
In this study, enzyme-powered millimotors 
have been made by crystallization 
of MOF particles 
encapsulating catalase enzymes 
onto soft cellulose microcapsules. 
The large size, but low mass of the 
motors makes them a unique addition 
to the range of existing 
particle motors as they are able to 
carry significant volumes of 
liquid cargo because of their 
capsule permeability. 
The large capsules are able to 
move using chemical fuel thanks to 
the numerous nanomotors 
attached to their surfaces, analogous 
to a large tanker ship driven by 
multiple tugboats.

Two mechanisms of propulsion 
result from enzyme-mediated 
production of  oxygen gas 
bubbles in the presence 
of hydrogen peroxide. 
Bubbles grow on and remain 
attached to hydrophobic 
motors, driving vertical 
motion as buoyancy rapidly grows.
For hydrophilic motors, 
bubbles are initially 
expelled from the structures, driving 
randomly-directed motion by 
recoil force, but as more 
bubbles grow motion 
becomes dominated by 
buoyancy and the 
motors rise similarly to the 
hydrophobic motors.
Despite their large size, the 
millimotors move with surprising 
speeds, reaching levels of 
80 motor diameters per second for 
the hydrophilic motors, when recoil 
force dominates movement, and 8 
motor diameters per second 
when buoyancy dominates.
The low density of the 
millimotors makes them 
highly efficient at movement, 
exhibiting similar or 
faster relative velocities 
than much smaller solid particle motors.
The flexible response to 
mechanical stress 
of the millimotors enables them to squeeze 
through significant 
constrictions, making their 
mobility viable even in 
confined spaces and environments 
where solid motor particles 
would clog and jam. 
The millimotors can compress their 
diameter by as much as 30\% in 
a constriction and their 
continuous generation of gas 
bubbles creates 
a low-slip 
surface layer and a back-pressure 
that enhances 
movement through 
tight spots. 
Such flexibility also 
offers a possible way to rapidly release 
the liquid cargo of the capsules 
and we plan to study this 
 in future work.

 \section*{Acknowledgement}
MH acknowledges support 
from a UNSW Scientia PhD 
fellowship. The authors 
acknowledge partial support 
from the Australian 
Government through the 
Australian Research 
Council's Discovery 
Projects funding scheme, 
Project DP190102614.
Lightsheet, confocal, 
and scanning electron 
microscopy were performed 
using instruments 
situated in, and maintained 
by, the Katharina Gaus 
Light Microscopy Facility 
(KGLMF) and Electron 
Microscope Unit (EMU) 
at the Mark Wainwright 
Analytical Centre, UNSW Sydney.


\appendix

\bibliography{mybibfile}

\end{document}